\documentclass[aps,prd,amsmath,twocolumn,longbibliography]{revtex4-1}

\usepackage{graphicx}\usepackage{multirow,hyperref,slashed}

\usepackage{amssymb,color}
\renewcommand{\bar}[1]{\overline{#1}}
\renewcommand{\bar}[1]{\overline{#1}}

\def\O{\Omega}
\def\e{\epsilon}
\def\Dslash{\raise.15ex\hbox{/}\kern-.7em D}
\def\Pslash{\raise.15ex\hbox{/}\kern-.7em P}

\def\s{\sigma}

\def\t{\tau}

\newcommand{\ben}{\begin{displaymath}}
\newcommand{\een}{\end{displaymath}}
\newcommand{\be}{\begin{equation}}
\newcommand{\ee}{\end{equation}}
\newcommand{\bwt}{\begin{widetext}}
\newcommand{\ewt}{\end{widetext}}
\newcommand{\bea}{\begin{eqnarray}}
\newcommand{\eea}{\end{eqnarray}}

\newcommand{\eq}[1]{Eq.~(\ref{#1})}

     \newcommand{\bpi}{\mbox{\boldmath$\pi$}}
    \newcommand{\bfpi}{\mbox{\boldmath$\pi$}}

\newcommand {\bft}{\mbox{\boldmath$\tau$}}

\def\g{\gamma}\def\a{\alpha}\def\d{\delta}

\def\m{\mu}

\newcommand{\beqn}{\begin{equation}}
\newcommand{\eeqn}{\end{equation}}

 \def\D{\Delta}

\def\bfa{{\bf a}}

\begin{document}
\title{The Fundamental Nature of Low-Energy Pion-Nucleon Interactions}
\author{Gerald A. Miller} 
\affiliation{ 
Department of Physics,
University of Washington, Seattle, WA 98195-1560, USA}

\date{\today}        
 
\begin{abstract} A historical review of pion-nucleon interactions at low energy is presented, with   aims  toward   an introductory,  pedagogic approach focusing  on issues germane to current research. These  topics include  the use of chiral effective field theory, the sigma term,  qualitative understanding of low-energy phase shifts, the physics of the $\Delta$ resonance,  and isospin violation. A theme is that low-energy pion-nucleon scattering is directly related to QCD via effective field theory. Another theme is that modern needs regarding the pion-nucleon sigma term and the study of isospin violation requires a better low-energy data set.
\end{abstract}

\maketitle     
\noindent

\section{Introduction} 

\newcommand{\beq}{\begin{equation}}        
\newcommand{\eeq}{\end{equation}}          
\def\g{\gamma}\def\m{\mu}
\def\a{\alpha}\def\d{\delta}\def\s{\sigma}\def\t{\tau}\def\O{\Omega}

The pion is the lightest of the strongly interacting particles, with a mass only  about 1/7 of that of the  nucleon.  The large 
Compton wavelength  of $1/m_\pi\approx 1.4 $ fm provides the longest-ranged
contribution to the nucleon-nucleon interaction and is therefore an  an element of all models, from
the ancient to the newest, of the nucleon- nucleon interaction. The one pion exchange potential is crucially responsible for the binding of nuclei~\cite{Kaiser:1997mw,Kaiser:2001jx}.
 Moreover, the presence of the pion as a significant component of the nuclear wave function is reinforced by the dominance of the pion in meson exchange corrections to a variety of nuclear properties. This was discussed long ago~\cite{Riska:1970jxh,Riska:1972zz,Brown:1973qgs} and recently~\cite{King:2020wmp}.   Moreover, the pion cloud of the nucleon presents an important component of the nucleon sea by causing the asymmetry of the nucleon sea, $\bar d>\bar u$, \cite{Alberg:2017ijg,Alberg:2021nmu} as verified by the SeaQuest experiment \cite{SeaQuest:2021zxb}.

The smallness of the pion mass is a central feature of theories that describe low-energy hadron physics. Knowledge of the mass hints to  an approximate symmetry of nature, chiral symmetry. A chiral transformation, in its simplest form,  changes a  fermion field-operator $\psi$ to $\g_5 \psi$. The Lagrangians of QED and QCD are invariant under this transformation except for terms involving mass, $m\bar \psi \psi$, and the small quark masses of QCD are related to the small value of the pion mass~\cite{GellMann:1968rz}.

The remainder of this review is mainly concerned with describing the consequences of the small pion mass for pion nucleon scattering. Sect. II is con- cerned with the fundamental Lagrangian as expressed in both quark and hadronic degrees of freedom. Sect. III describes chiral symmetry breaking and the sigma term. The current importance of the sigma term and means to extract it are also described. Sec. IV de- scribes the low-energy phase shifts. A pedagogic de- scription of the $\Delta$  resonance is presented in Sect. V. Isospin violation is described in Sect. VI. A brief summary appears in Sect. VII.
\section{The Chiral Lagrangian and its Relation to QCD}
The well-known QCD Lagrangian for the light up and down quark flavors is given by
\bea&
   {\cal L}(x)=\bar q(x)\gamma^\mu\left(i\partial_\mu -g {\lambda^a\over 2}G^a_\mu\right)q(x)\nonumber\\& -\bar q(x)m_qq(x)-{1\over 4} G^{a\mu\nu}G^a_{\mu\nu} \>,
   \label{QCD}
\eea
where $G^a_\mu$ are the gluon-field operators, $a$ represents the color, $q$ are the quark-field operators (a column vector),  $\lambda^a$ are the eight matrices generating $SU(3)_{\rm color}$  and
\begin{equation}
    m_q=\begin{bmatrix}
    m_u &0\\
    0& m_d
    \end{bmatrix}
\label{mm}\end{equation}
is the quark-mass matrix. The masses in the Lagrangian are known as current quark masses. If $m_d=m_u$ the theory is invariant under isospin rotations:
\begin{equation}
    q(x)\rightarrow e^{i\bfa\cdot\bft/2}q(x) \>,
\end{equation}
where $\bfa$ is a constant vector. The relevant transformation that interchanges the $u$ and $d$ quarks is obtained with $\bfa=\pi {\bf \hat y}$, a rotation of $\pi$ about the $y-$ axis in isospin space, with the $z-$ axis associated with charge. This transformation is known as charge symmetry, a specific rotation in isospin space~\cite{Miller:1990iz,Miller:2006tv}.

In the further limit that the current quark masses vanish $m_u=m_d=0$ the right- and left handed $(1\pm \g_5)q$ decouple and the theory  is symmetric under two copies of isospin symmetry:
\beq
    q(x)\rightarrow e^{i\bfa\cdot\bft(1+\g_5)/2}q(x) ,\quad  q(x)\rightarrow e^{i\bfa\cdot\bft(1-\g_5)/2}q(x)
 \>.
 \label{chiral}
 \eeq
These chiral transformations  represent the global symmetry group SU(2)$_R\times$SU(2)$_L$. The current quark masses are small enough so that QCD respects this symmetry to a good approximation.

The conserved Noether currents related to the transformations of \eq{chiral}  are the isospin currents:
\beq  
    J_\mu^a(x)=\bar q(x)\g_\mu{\t^a\over 2}q(x)
\eeq
and the axial currents
\beq  
    A^a_\m(x)=\bar q(x)\g_\mu\g_5{\t^a\over 2}q(x) \>.
\eeq
The  volume integral of $A_0^a$ is known as the axial charge $Q_A^a$.

The elegant QCD Lagrangian is not immediately applicable to 
 understanding low-energy pion-nucleon scattering because of the non-perturbative  difficulties associated with strong interactions at low energies. However there is another long-standing technique: that of using effective Lagrangians that embody the physics of QCD.  The parameters of the effective Lagrangian can sometimes  be computed using   lattice QCD techniques, and this aspect is discussed below.

Weinberg~\cite{Weinberg:1978kz} wrote that quantum field theory has no content beyond analyticity, unitarity, cluster decomposition and symmetry. In particular, 
``if one writes the most general possible Lagrangian, including {\it all} terms consistent with assumed symmetry principles, and then computes matrix elements with this Lagrangian to any given order of perturbation theory, the result will be the most general $S-$matrix consistent with analyticity, perturbative unitarity, cluster decomposition and the assumed symmetry principles.".

This means that a properly constructed theory of pion-nucleon interactions contains, in principle, the same information content as QCD. Here the symmetries are isospin and chiral. Although approximate, using these should give a good guide to the correct results. 
Weinberg~\cite{Weinberg:1994tu} illustrates the suggested procedure by considering a theory of massless pions, governed by a chiral SU(2)$_R\times$SU(2)$_L$ symmetry. The Lagrangian will be 
SU(2)$_R\times$SU(2)$_L$ invariant provided it conserves isospin and is constructed only from a chiral-covariant derivative of the pion field, which may be taken to be
\beq 
    { D}_\mu\bpi= {\partial _\mu\bpi \over 1+\bpi^2/4f_\pi^2} .
\eeq
Then the most general such Lagrangian consists of an infinite series of operators:
\beq 
    {\cal L}=\sum_{n=1} g_n ({ D}_\mu\bpi\cdot { D}_\mu\bpi)^n .
\eeq
When this Lagrangian is used to compute pionic $S$-matrix elements it yields the most general matrix elements consistent with the stated general principles, provided that all terms of all orders in the coupling constants $g_n$ are included.
Weinberg explains  that using the effective Lagrangian allows a systematic way to calculate corrections to results of leading order.
One may derive an expansion in powers of (small values) of  momentum divided by a large mass scale.

The Lagrangian for pion-baryon  interactions  is more complicated.  Displaying only the interaction terms to the first-and second- order in powers of the pion field, we use \def\e{\epsilon}
\bea&
    \mathcal{L}_{\rm int} = 
    - \; {g_A\over 2 f_\pi} \bar\psi \gamma_\mu \gamma_5 \tau^a \psi 
    \, \partial_\mu \pi^a 
    - {1\over 4f_\pi^2} \bar\psi \gamma_\mu \tau^a \psi \;
    \e^{abc} \pi^b \partial_\mu \pi^c\nonumber\\& -{g_{\pi N\Delta}\over 2M} (\bar{\Delta}^i_\mu g^{\mu\nu}\psi\partial_\nu \pi^i +{\rm H. C.}) \>,
\label{lag}
\eea
where $\psi$ is the nucleon  Dirac field  of the nucleon,   $\pi^a (a = 1, 2, 3)$ the chiral pion field and $M$ is the nucleon mass. In Eq.~(\ref{lag}) $g_A$ denotes the nucleon axial vector coupling and $f_\pi$ the pion decay constant.  The second  term  is the Weinberg-Tomozawa term  which describes  
low-energy $\pi-$nucleon scattering. In the third term $g_{\pi N\Delta}$ is the $\pi N\Delta$ coupling constant, and the $\Delta^i_\mu$ field   is a vector   in both spin and  isospin space. Notice that only derivatives of the pion field enter. 

The meson factories LAMPF, SIN (now PSI) and TRIUMF began to operate in the 1907's with a focus on pion-nucleon and pion-nucleus interactions with an earlier motivation involving the pion as a fundamental carrier of the strong interaction. The very close connection between the hadronic Lagrangians in use at the time and QCD was mainly  not understood during the peak of the activities of the meson factories~\cite{ericson1991}. Thus opportunities for improved studies of the fundamental interaction involving better low-energy pion  beams were not realized or perhaps not  even dreamed about. We shall see that determination of the pion-nucleon sigma term would be greatly improved by better quality data. 

The  Lagrangian of \eq{lag} enjoys remarkable success when applied to compute pion-nucleon scattering amplitudes at threshold. Then terms  involving spatial momenta drop out. A simple discussion is provided in the book by Ericson \& Weise~\cite{Ericson:1988gk}.   The only surviving term, expressed as a Hamiltonian  density ${\cal H}$ is given by
\beq 
    {\cal H}= {1\over 4f_\pi^2}\bft\cdot (\bpi \times {\partial \bpi\over \partial t}) \>.
\eeq

This  interaction may be used to compute the isovector $s$-wave pion-nucleon interaction. First take the matrix elements between in- and outgoing pion states to find
\beq
    \langle \pi_b|\bft\cdot (\bpi \times {\partial \bpi\over \partial t})|\pi_b\rangle= 2i m_\pi \e_{abc}\tau_c \>.
\eeq
This Hamiltonian density  corresponds to a zero range (s-wave) pion-nucleon potential, $V_s({\bf r}) $ given by
\beq
    V_s({\bf r})={1\over 4f_\pi^2}\bft\cdot {\bf t}\delta({\bf r}) \>,
\label{delta}\eeq
where $\bf t$ is the pion isospin operator. Using the Born approximation  the scattering length, $a$ is given by 
\beq 
    a=-{\tilde m_\pi} \int d^3r V_s({\bf r}) \>,
\eeq
where $\tilde m_\pi=m_\pi/(1+m_\pi/M)$ is the pion-nucleon reduced mass.
The isospin operator operator $\bft\cdot {\bf t}$ evaluates to  1 for an $\pi$N isospin state $3/2$ and $-2$ for an isospin state of 1/2. Thus the two scattering lengths $a_{1,3}$ are given by
\beq 
    a_1 = {\tilde m_\pi\over 4\pi f_\pi^2 },\quad a_3={-\tilde m_\pi\over 8\pi f_\pi^2 } \>.
\eeq 
The scattering length can be expressed 
as
\beq 
    a = (a_1 + 2 a_3)/3+ (a_3-a_1)/3\,\bft\cdot {\bf t} \label{adef}
\eeq 
so the isovector combination is  
\beq 
    a_1 - a_3 = 3{{\tilde m_\pi}\over  8\pi f_\pi^2}= 0.234 m_\pi^{-1}  
\label{est}\eeq 
and the isoscalar term is given by
\beq
    a_1 + 2a_3 = 0 \>.
\eeq  

A more complete analysis is necessary because the second Born approximation for the potential of \eq{delta} is divergent. This has been provided via the detailed chiral perturbation theory analyses~\cite{Fettes:1998ud,Fettes:1998wf,Fettes:2000gb,Fettes:2000bb,Fettes:2000xg,Fettes:2000vm,Fettes:2001cr} that
~\cite{Fettes:1998ud} finds 
 \beq 
    a_1-a_3=  0.264(15)  m_\pi^{-1},\label{emp}
\eeq
while~\cite{Ericson:1988gk} reports 
\beq
 a_1+2a_3=-0.029(9) m_\pi^{-1} .
 \eeq
The large magnitude of the isospin-dependent term and the small magnitude of the isospin-independent term of the scattering length, $a$,   are a striking success of the simple Lagrangian.  It has been argued~\cite{Weinberg,Long:2009wq}, in the spirit of an expansion in the pion mass,  that it is the pion mass, instead of the reduced pion mass, that should appear in the above equations. In that case the agreement between the simple estimate \eq{est} and empirical extraction~\eq{emp} would be perfect.

\section{Chiral Symmetry Breaking and the Sigma Term}

The quark mass term of \eq{QCD} is not invariant under the transformations of \eq{chiral}. It may be put in the form~\cite{Weinberg:1994tu}:
\beq 
    {\cal L}_{\rm mass}=-(m_u+m_d)V_4-(m_u-m_d)A_3 \>,\label{mud}
\eeq
where
\beq 
    V_4={1\over 2}(\bar u u+\bar d d ),\quad A_3={1\over 2}(\bar u  u-\bar d d  \>).
\eeq
The operators $V_4$ and $A_3$ are spatial scalars and components of independent chiral four vectors $V_\a$ and $A_\a$. Following Weinberg, terms must be added to the effective chiral Lagrangian with the same transformation properties. The aim is to construct chiral symmetry breaking terms without derivatives of the pion field. Using only the pion field, there is no Lagrangian term in $A_3$ and only  one term in $V_4$. This is the pion mass term
\beq  
    V_4(m_\pi)=-{m_\pi^2\pi^2\over 2(1+\pi^2/(4f_\pi^2))} \>.
\eeq  
Using pion fields and a nucleon bi-linear (with no derivatives) and assuming the isospin limit of $m_u=m_d$ one finds a term
\beq  
    \d{\cal L}_{\rm int}=-{A\over 2}\left({1-\pi^2/4f_\pi^2\over 1+\pi^2/4f_\pi^2}\right) \bar\psi\,\psi, \label{At}
\eeq 
where $A$ is a constant proportional to $m_u+m_d$ and $\psi$ is the nucleon doublet. This term gives a contribution to the isoscalar scattering length (the first term of \eq{adef}):
\beq  
    {a_1+2a_3\over 3}={A\over 2f_\pi^2} \>.
\eeq

The term $A$ is the sigma  term, $\s_{\pi N}$, which has recently emerged as a quantity related to many aspects of physics, such as the origin of the mass of ordinary matter, studies of dark matter,  nuclear binding and nucleosynthesis.
The plan here is to first consider the relation of  $\s_{\pi N}$ to pion-nucleon scattering and then discuss its significance. My discussion parallels that of the  recent  review   by Alarc\'on~\cite{Alarcon:2021dlz}.


%


Consider the pion-nucleon scattering process:
\bea
 \pi ^a (q) N(p) \to  \pi^b (q') N(p'). \label{pin}
\eea
where $a\ (b)$ are the isospin indices of the initial (final) pion of momentum $q\ (q')$, and $p\ (p')$ the momentum of the initial (final) nucleon.
The off-shell scattering  amplitude  for pions of squared four-momentum $q^2$ and $q'^2$, $T^{ba}$, is given by \cite{Reya:1974gk}
\begin{widetext}
\bea\label{Eq:T2}&
{ T^{ba}(\nu, t, q^2, q'^2) \over K(q,q')}= \int d^4x \langle N(p')| q'_\mu q_\gamma T(  A^{b \mu}(x)  \partial_\gamma A^{a\gamma}(0))+ i q'_\mu \delta(x_0) [A^{b \mu}(x) ,\partial_\gamma  A^{a\gamma}(0)] |N(p) \rangle e^{i q' \cdot x} \nonumber \\&
   - \int d^4x \langle N(p')|\delta(x_0) [A^{b 0}(x) , A^{a\gamma}(0)]  |N(p) \rangle e^{i q' \cdot x}. 
\eea
\end{widetext}
where $K(q,q')\equiv i \frac{(q^2 - m_\pi^2)(q'^2 - m_\pi^2)}{m_\pi^4 f_\pi^2},\,\nu = \frac{s - u}{4 M}$, with  $M$ the nucleon mass, $A^{a\mu}$ the axial current, $f_\pi$ the weak decay constant of the pion and $m_\pi$ the pion mass.
 The last term is not determined by the commutation relations  provided  by current algebra, but instead is isolated by taking the so-called ``soft-meson limit" $q'_\mu, q_\nu \to 0$. Then,
 using  Eq.~\eqref{Eq:T2}  and $\partial_\nu A^{a\nu}=-i [Q_5^a,\mathcal{H}  ]$ yields
\bea&\label{Eq:Def-sigma-commutator1}
 T^{ba}(0, 0, 0, 0) 
                            &=  - \frac{1}{f_\pi^2} \langle N(p')| [Q_A^{b}(0) , [Q_A^{a}(0), \mathcal{H}(0) ]]|N(p) \rangle\nonumber\\&
\eea 
where  $Q_A^i$ are the axial charges and $ \mathcal{H}$ is  the Hamiltonian density. Only terms  that break  chiral symmetry enter in its commutator with the axial charges. 
The double commutator of the last matrix element of Eq.~\eqref{Eq:Def-sigma-commutator1} is the  $\pi N$ sigma commutator, $\sigma_{\pi N}$. 
The Jacobi identity tells us that the sigma commutator is symmetric in the isospin indices  and, therefore, contributes only to the isoscalar part of the scattering amplitude. The mass  matrix $m_q$ of \eq{mm}
  provides the part of  $\mathcal{H}$ that does not commute with the axial charge and yields
\bea
\sigma_{\pi N} =\frac{\bar{m}}{2 M} \langle N(p)| \bar{u}u + \bar{d}d  |N(p) \rangle,
\eea 
where  $\bar m \equiv \frac{m_u+ m_d}{2}$,
usually called the {pion-nucleon sigma term}. The  factor $2 M$ in the denominator arises from the covariant normalization $\langle N(p')|  N(p)\rangle = (2\pi)^3 2E\, \delta({\vec{p}\,' - \vec{p}})$, used here. Note also that $\bar{m}$ arises because here we use $m_d=m_u$. 

The sigma term  appears in many areas of modern physics as explained below. First it is necessary to understand its rather  indirect relation to scattering. 
The amplitude of Eq.(\ref{Eq:Def-sigma-commutator1})  is not evaluated at  measurable  kinematics, so one needs to find a way to connect the experimental data  to  $T^{ba}(0, 0, 0, 0)$. 
 Ref.~\cite{Cheng:1970mx} showed that  one can expand the amplitude in $q^2$ and $q'^2$ to relate the physical amplitude to its soft  limit. One 
 invokes the Adler consistency conditions \cite{Adler} and evaluates the amplitude at $(\nu = 0, t = 2m_\pi^2)$  to cancel the leading correction in $q^2$ and $q'^2$, and relate the physical amplitude to $T^{ba}(0, 0, 0, 0)$ plus a remainder. 
This kinematical point is called the Cheng-Dashen point (with $\nu=0,\,s=m_N^2, t=2m_\pi^2$)  and lies in the unphysical (subthreshold) region of $\pi N$ scattering. The ability to extract the amplitudes depends on having very high quality scattering data at low energies and a very precise method of handling the effects of the Coulomb interaction. The date begins for pion kinetic energies of about only 30 MeV. 
Moreover, there is no unique way to split the interaction into strong and electromagnetic parts~\cite{Gasser:2003hk}, so in general models must be used.

Cheng and Dashen,  assumed that the remainder was of order $m_\pi^4$, but 
Brown, Pardee and Peccei showed that there were non-analytic contributions that made this relation valid only up to order $m_\pi^2$ \cite{Brown:1971pn}. 
They presented a widely-used  modified form of the Cheng-Dashen theorem  valid up to $\mathcal{O}(m_\pi^4)$.
To be more specific, one writes the $\pi N$ scattering amplitude in terms of the four Lorentz invariant amplitudes $A^\pm$, $B^\pm$
\bea\label{Eq:T-decomposition}
 &T^{ba} = \delta^{ba} T^+ + \frac{1}{2}\left[ \tau^b, \tau^a\right] T^- \\
 &T^\pm  = \bar{u}' \left[  A^\pm + \frac{1}{2} (\slashed{q}' + \slashed{q} )B^\pm \right] u  \eea
where $ T^\pm$ are the isoscalar and isovector parts of the amplitude. The combination $D^\pm \equiv A^\pm + \nu B^\pm$ is amenable to a detailed analysis. The original result by Cheng and Dashen states that $\Sigma_{\pi N}\equiv f_\pi^2  \bar{D}^+(\nu = 0, t = 2 m_\pi^2)$ and $\sigma_{\pi N}$ are equal up to order $m_\pi^2$, with $\bar{D}^+$ the Born-subtracted part of the amplitude.
In Ref.~\cite{Brown:1971pn} this theorem was recast as follows
\bea&\label{Eq:Pagels&Pardee}
 \Sigma_{\pi N}=f_\pi^2  \bar{D}^+(\nu = 0, t = 2 m_\pi^2) \nonumber\\&= \sigma(t = 2 m_\pi^2) + \Delta_R,
\eea
where 
\bea
\sigma(t) = \frac{1}{2 M} \langle N(p')| \hat{m}(\bar{u}u + \bar{d}d)  |N(p) \rangle.
\eea

This expression is useful because   $\Delta_R$ is of order $m_\pi^4$ and small ($\Delta_R \simeq 2$~MeV \cite{Bernard:1996nu}), and 
the scalar form factor of the nucleon at $t = 2m_\pi^2$, $\sigma(t = 2m_\pi^2)$, can be related to the sigma term though dispersion relations or other methods, because  $\sigma(t=0) = \sigma_{\pi N}$.
The latter was first calculated in Ref.~\cite{Pagels:1972kh}. 


\subsection{Why is the sigma term important?}
\label{Sec2}

This sub-section presents a brief discussion explaining the current importance of the sigma term.

\subsubsection{Origin of the mass of ordinary matter}

Recently  there has been an increasing effort to understand the  the proton (or nucleon) mass in terms of QCD, 
see {\textit e.g.} ~\cite{Ji:1994av,He:1994gz,Metz:2020vxd,Lorce:2021xku,Ji:2021qgo}. 
The usual strategy to analyze its decomposition is to consider that the energy-momentum tensor of QCD can be related to the mass of a hadron (in this case the nucleon) in the following way: 
\bea\label{Eq:nucleon-mass-1}
 M= \frac{1}{2M}\langle N(p) | T^\mu_{\ \mu} | N(p) \rangle 
\eea 
where $T^\mu_{\ \mu}$ is the trace of the energy momentum tensor.
This trace, in its renormalized form, is given by \cite{Crewther:1972kn,Chanowitz:1972vd,Collins:1976yq,Shifman:1978zn}:  
\bea\label{Eq:nucleon-mass-2}
 &M 
 = \frac{1}{2M}\langle N(p) | \frac{\beta(\alpha_s)}{2\alpha_s} G^{a \mu \eta }G^{a}_{\ \mu \eta}\nonumber \\
 &+ \sum_{q = u, d, s} \gamma_{m_q} m_q \bar{q} q   | N(p) \rangle +  \frac{1}{2M}\langle N(p) | \sum_{q } m_q\bar{q} q  | N(p) \rangle,\nonumber\\&
\eea
where  
$\alpha_s$ is the strong coupling constant, $\beta(\alpha_s)$ its beta function, $m_q$ is the mass of the quark $q$ and $\gamma_{m_q}$ the anomalous dimension of the mass operator.
In the limit in which strange quarks are neglected, the last term in \eqref{Eq:nucleon-mass-2} is $\sigma_{\pi N}$.
This term is   renormalization scheme and scale independent, and can be interpreted as the light quark mass contribution to the nucleon mass. This term arises from the Higgs boson couplings to quarks, so that
 its  accurate determination   would tell us the amount   of ordinary matter mass that is  generated though this mechanism and how much is dynamically generated. Other mass decompositions include \cite{Ji:1994av,He:1994gz,Metz:2020vxd,Lorce:2021xku,Ji:2021qgo}, but all contain the quark-mass  term.


\subsubsection{Dark matter detection}

The main recent  interest in the  precise value  of $\sigma_{\pi N}$ arises from its role in for detecting scalar dark matter. If one assumes that
 dark matter is a scalar, an effective Lagrangian that couples dark matter ($\chi$) to quarks ($q$) is given by:
\bea
 \mathcal{L}_{\chi q} = C \frac{m_q}{\Lambda^3} \bar{\chi} \chi \bar{q}q
\eea
where $C$ is a Wilson coefficient  containing information regarding  energy scales higher than the cutoff $\Lambda$. 

This effective Lagrangian leads to a  $\chi-$ nucleon  and consequently a  $\chi-$nuclear cross section \cite{Hill:2011be} that is proportional to the square of $\sigma_{\pi N}$ (if $u$ and $d$ quarks are sufficient) so that 
uncertainties in $\sigma_{\pi N}$ are  magnified in dark matter searches  \cite{Bottino,Ellis:2008hf}.These authors stressed the vital  importance of an  experimental campaign to determine a better value of  $\sigma_{\pi N}$.

It is noteworthy that Ref.~\cite{Crivellin:2013ipa}  showed how to avoid unnecessary and uncontrolled assumptions usually made in the literature about soft SU(3) flavor symmetry breaking in determining the two-flavor nucleon matrix elements relevant for direct detection of weakly interacting massive particles (WIMPs).   This work was the first to allowfor an accurate assessment of hadronic uncertainties in spin-independent WIMP-nucleon scattering and for a reliable calculation of isospin-violating effects.


\subsubsection{Nuclear binding and Nucleosynthesis }

The $\sigma$-model Lagrangian~\cite{Gell-Mann:1960mvl,Brown:1971pn,Noble:1979iv}  contains a term that violates chiral symmetry of the form  $(m^2_\s-m_\pi^2)/(2M) g \s\bfpi\cdot\bfpi$,
where $\s$ represents the field of a scalar-isoscalar interaction and $g$ is the pion-nucleon coupling constant. This terms yields a contribution to $T^+$ of \eq{Eq:T-decomposition} and therefore to $\s_{\pi N}$. The $\s$-model when enhanced by the inclusion of vector mesons incorporates the  dynamics necessary to bind nuclei: long- and medium-ranged attractive forces and short-ranged repulsive forces.

More generally,
the pion-nucleon sigma term is related to the strength of the $\pi N$ interaction with scalar-isoscalar quantum numbers. 
This part of the interaction plays a prominent role in alpha-like particle clustering \cite{Elhatisari:2016owd}.
Understanding this process becomes essential in order to explain the observed ${}^{12}\text{C}$ abundance, which is possible only though to the formation of a $0^+$ resonance near the ${}^4\text{He }\text{+ }{}^8\text{Be}$ threshold (the Hoyle state). This state is important for nucleosynthesis.

\subsubsection{Nuclear thermodynamics}

The matrix element of QCD related to the sigma term also provides valuable information about the properties of nuclear matter. 
When one studies the restoration of chiral symmetry in nuclear matter, one of the quantities used as order parameter in the phase transition is the chiral condensate in the nuclear medium, $\langle \Psi | \bar{q}q |\Psi \rangle$, where $|\Psi\rangle$ is the nuclear ground state \cite{Weise:2012yv}. 
Applying the Hellmann-Feynman theorem,for nuclear matter of Fermi momentum $k_F$, to the Hamiltonian density of QCD one finds
  \cite{Kaiser:2007nv}
  \begin{widetext}
\bea\label{Eq:Condensate}
 \frac{\langle \Psi | \bar{q}q |\Psi \rangle}{\langle 0 | \bar{q}q | 0 \rangle}(\rho) = 1 - \frac{\rho}{f_\pi^2} \left\{ \frac{\sigma_{\pi N}}{m_\pi^2} \left( 1- \frac{3k_f^2}{10 m_N^2} + \frac{9 k_f^4}{56 m_N^4} \right) + {\partial E(\rho)/A\over \partial m_\pi^2}  \right\} 
\eea
\end{widetext}
where $ \langle 0 |\bar{q}q | 0 \rangle$ is the quark condensate in the vacuum, 
and $E$ is the energy of nuclear matter.  
This means  that the sigma term is important for determining the density for  the transition to a chiral symmetric regime in which $ \langle \Psi | \bar{q}q |\Psi \rangle=0.$
Increasing  the value of   $\sigma_{\pi N}$ reduces the value of the transition density. However, the interaction terms $ {\partial E(\rho)/A\over \partial m_\pi^2}  $ are very important  \cite{Kaiser:2007nv}.

\subsubsection{Strangeness content of the nucleon}

The quantity $\langle N | m_s \bar{s}s| N \rangle\equiv 2M\sigma_s$ is related to the contribution of the strange quark to the nucleon mass. 
This matrix element has been  related to $\sigma_{\pi N}$ though group-theoretical arguments as follows.
Consider
 the extended  mass term of the QCD Hamiltonian:
\bea&
 \mathcal{H}_m    
  = \frac{1}{3}(m_s + 2 \bar{m})( \bar{u}u + \bar{d}d + \bar{s}s ) \nonumber\\&+ \frac{1}{3}(\bar{m} - m_s)( \bar{u}u + \bar{d}d -2 \bar{s}s ).
\eea
The second term  can be related to the mass splitting of the strange and non-strange baryons at leading order in the $SU(3)$-breaking parameter $m_s- \bar{m}$, using octet mass-splitting \cite{Alarcon:2012nr,Fernando:2018jrz}:
\bea&\label{Eq:sigma0}
 \sigma_0 \equiv \frac{\bar{m}}{2 M} \langle N |( \bar{u}u + \bar{d}d -2 \bar{s}s )|N \rangle\nonumber\\& = \frac{\bar{m}}{m_s - \bar{m}} (m_\Xi + m_\Sigma - 2 M).
\eea

The relation of $\sigma_0$ with the octet baryon mass splitting provides an estimate of $\sigma_s$, assuming a known value of $\sigma_{\pi N}$:
\bea
 \sigma_s = \frac{m_s}{2\hat{m}} (\sigma_{\pi N} - \sigma_0)
\eea
Assuming  $\langle N | \bar{s}s | N \rangle=0$ (Zweig rule), the value of $\sigma_{\pi n}$ should be well approximated by $\sigma_0$  \cite{Cheng:1975wm}. 
Then  using the present day-values of neutral baryon masses and the
current quark ratio~\cite{Zyla:2020zbs} :  $m_s/2\bar{m}\approx 27$. then $\sigma_{\pi n}$   should be around $24$~MeV. Estimates of the SU(3)-breaking effects
increase this value to about 35 MeV \cite{Gasser:1980sb}. If one allows for a non-zero value of $\sigma_s$, then  one finds
$\sigma_{\pi N}\approx 0.073\, \sigma_s+35\, \rm MeV.
$
Then using a typical value of  $\sigma_{\pi N} \approx  60$~MeV one obtains    $\sigma_s\approx 350 $ MeV,  a huge value. In contrast, lattice QCD calculations give results for $\s_s$ between 30 and 50 MeV 
\cite{Durr:2011mp,Durr:2015dna,Yang:2015uis,Freeman:2012ry,Junnarkar:2013ac}.
 Recent developments, discussed in the next section  show that  $\s_{\pi N}$ can be about 60 MeV without a need for a huge value of $\s_s$. Therefore the above analysis is  seems dated. 
 
The strangeness content of the nucleon, related to $\s_s$ has also been postulated to reduce the  proton spin  below the value obtained in simple quark models, and therefore play a role in the proton spin crisis;  see the reviews \cite{Hughes:1999wr,Beck:2001yx}.

\subsection{Methods to determine  the sigma term}

In this section I    briefly discuss  some of the different approaches and strategies that have been used to determine the sigma term. The problem in using experimental measurements  is in relating
$T(0,0,,0,0)$ to  measurable kinematics in which $q^2={q'}^2=m_\pi^2$.

The earliest attempts used the simple assumption, explained above that using Eq. \ref{Eq:sigma0} with $\sigma_s=0$ gives $\sigma_{\pi N}\approx 35$ MeV.  The second is the use of dispersion relations that allow a continuation from $\pi N$ scattering data to the Cheng-Dashen point See the review 
 \cite{Alarcon:2021dlz} and the original works \cite{Gasser:1988jt,Gasser:1990ap,Gasser:1990ce,Bernard:1996nu,Hoferichter:2015dsa}.
Sum rules can also be used to express $\Sigma$ in terms of pion nucleon threshold parameters  \cite{Altarelli:1971kh,Olsson:1979ee,Olsson:1999jt}

Another method is the use of chiral perturbation theory.  As a systematic way to determine  QCD Green's functions \cite{Gasser&Leutwyler},
chiral EFT is widely used  to  study  strong interactions at low energies. 
Chiral EFT is constructed to systematically provide corrections to the low-energy theorems of QCD, so it is  a natural tool to calculate the pion-nucleon sigma term. One writes a Lagrangian that complies with chiral symmetry, with unknown short-distance, high-momentum transfer effects modeled by the use of low energy constants that are determined by data. Then   explicit chiral symmetry breaking is provided by including a term
of the form
\bea\label{Eq:LEC_c1}
 \mathcal{L}_{\pi N}^{(2)} 	\supset c_1 \langle \chi_+ \rangle \bar{\psi}\psi  
\eea
where $\psi$ refers to the nucleon isodoublet, and  $\langle \chi_+ \rangle$ is a trace in the isospin space of a matrix involving even powers of the pion field and quark masses (see Ref.~\cite{Alarcon:2012kn} for more details).  The value of $c_1$ is determined by using the Lagrangian to compute $\pi N$ phase shifts  fits to data, and accurate  results for $c_1$ are obtained only if all of the low energy constants can be determined.

Lattice QCD is well-known to  allows direct, non-perturbative numerical calculations using the QCD Lagrangiani.
a numerical method to compute QCD Green functions. 
 Two different strategies have been used.
The first one is to calculate the sigma term through the Hellmann-Feynman theorem. 
For that, one needs the quark mass dependence of the nucleon mass. Another approach is to calculate the sigma term as the scalar form factor of the nucleon at $q^2 = 0$~\cite{Durr:2015dna,Yang:2015uis,Abdel-Rehim:2016won,Bali:2016lvx,Alexandrou:2017qyt,Borsanyi:2020bpd}.

Ref.~\cite{Friedman:2019zhc}  used pionic $\pi H$ atom data to  estimate  the sigma term. 
The strategy of this approach is to relate the observed modification of the isovector scattering length $a_{0+}^-\equiv (a_1-a_3)/3$ in pionic atoms to $\sigma_{\pi N}$. 
They construct the relation from the leading order chiral result for $a_{0+}^-$ in vacuum, given by the Weinberg-Tomozawa term
\bea
a_{0+}^- = - \frac{M_\pi m_N}{8\pi(m_N + M_\pi)f_\pi^2 } .
\eea

One  obtains  the in-medium modification of $a_{0+}^-$ by noting that the ratio of condensates appearing in Eq.~\eqref{Eq:Condensate} is equal to (following~\cite{GellMann:1968rz})  the ratio $\frac{f_\pi^2(\rho)}{f_\pi^2}$
then 
\bea
\frac{f_\pi^2(\rho)}{f_\pi^2} = \frac{\langle \Psi | \bar{q}q |\Psi \rangle}{\langle 0 | \bar{q}q | 0 \rangle}(\rho) \approx 1 - \rho  \frac{\sigma_{\pi N}}{f_\pi^2 M_\pi^2} 
\eea
which modifies the free scattering length to 
\bea
 \hat{a}_{0+}^- = a_{0+}^- \left( 1- \rho\frac{\sigma_{\pi N}}{f_\pi^2 M_\pi^2 } \right)^{-1}
\eea
where $\hat{a}_{0+}^-$ refers to the in-medium value of $a_{0+}^-$.
For details see \cite{Friedman:2019zhc}.
This method is based on the medium modificaton of the dominant isovector s-wave term. But the  medium modification also modifies the  nuclear isoscalar s-wave term that is much larger than predicted by the impulse approximation  \cite{Miller:1980zza,Miller:1980zzb,Olivier:1984wm,Chakravarti:1993cy} of the pion nucleus optical potential related to the 
isoscalar s-wave term.  

\subsection{Determining the sigma term}

Alarcon \cite{Alarcon:2021dlz} presented a detailed discussion of the lengthy history of determinations  and  the great variation of the results. 
The latest results are tabulated in the 2021 FLAG review \cite{Aoki:2021kgd}. We present their results in Figure~1.

\begin{figure}[t]\label{FLAG}
 \begin{center}
 \includegraphics[width=.516\textwidth]{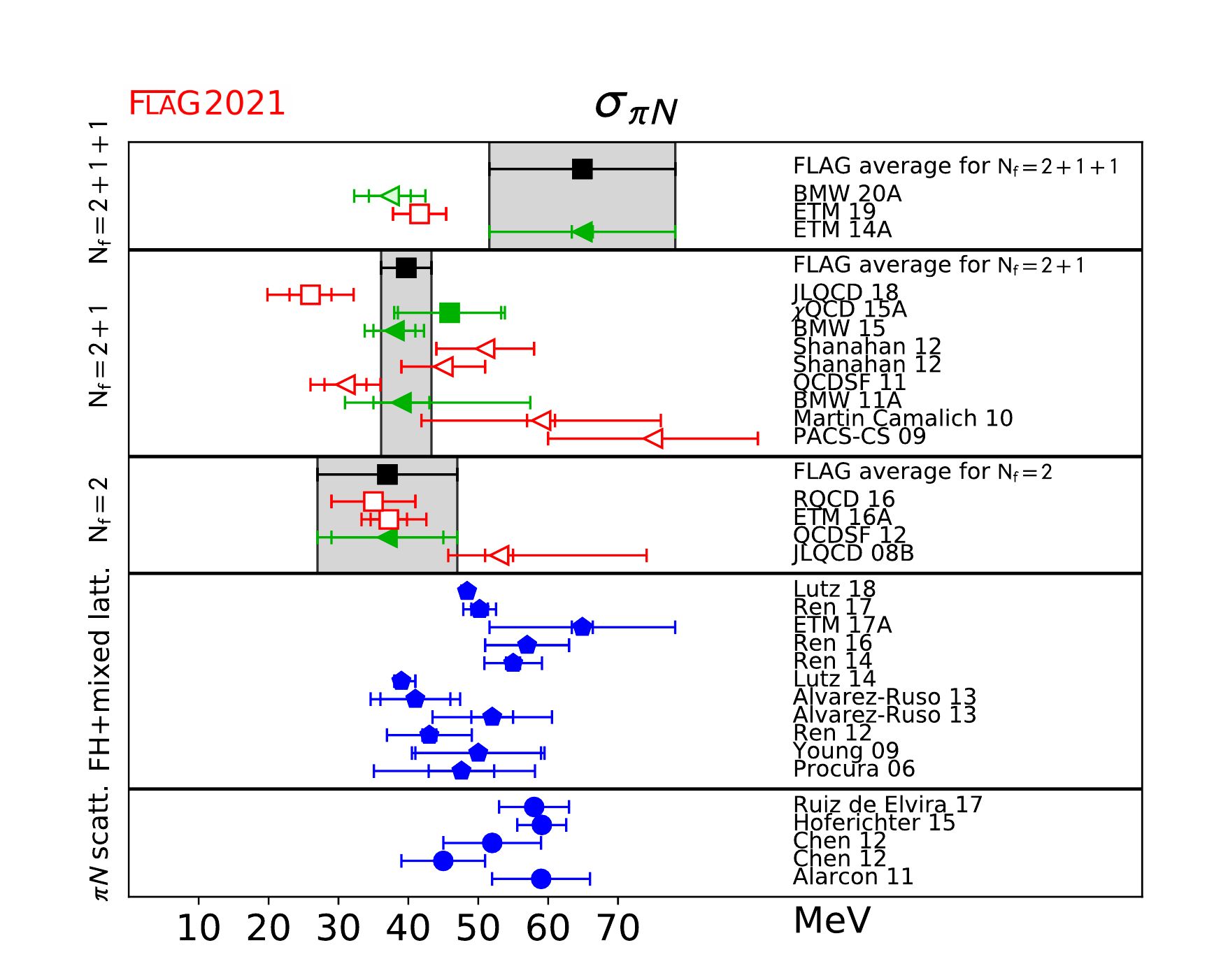} 
 \caption{ Lattice results and FLAG averages for the nucleon sigma term,$\sigma_{\pi N}$  for the $N_f = 2, 2 + 1$, and 2 + 1 + 1 flavor calculations. Determinations via the direct approach are indicated by squares and the Feynman-Hellmann method by triangles. Results from calculations which analyze more than one lattice data set within the Feynman-Hellmann approach \cite{Alexandrou:2017xwd,Procura:2006bj,Young:2009zb,Ren:2012aj,Alvarez-Ruso:2013fza,Lutz:2014oxa,Ren:2014vea,Ren:2016aeo,Ren:2017fbv,Lutz:2018cqo}, 
 are shown for comparison (pentagons) along with those from recent analyses of $\pi N$ scattering \cite{Alarcon:2011zs,Chen:2012nx,Hoferichter:2015dsa,RuizdeElvira:2017stg} 
 (circles). The results that pass the FLAG criteria are shown in green, and are reasonably consistent. }
 \end{center}
\end{figure} 

FLAG presents its results as
$ N_f= 2 + 1 + 1: \sigma_{\pi N} = 64.9(1.5)(13.2)$ MeV, a result that is larger than  for the case of $N_f=2+1$ of  $37(8)(6) $ MeV. 
The meaning of 2 + 1+ 1 is that $m_u=m_d<m_s<m_c$.The latter value  is believed to be more reliable because of the smaller error bar.  A weighted average of the two results would greatly favor the more precise latter value. 
The FLAG review  cite a tension with the result of \cite{Hoferichter:2015dsa} of $\sigma_{\pi N} =(59.1\pm 3.5)$ MeV of about three or four standard deviations. A significant source of uncertainty in lattice calculations is  contamination due to excited states. A recent, calculation \cite{Gupta:2021ahb}  forces the first excited state to  be that of  a nucleon-pion $ p$-wave energy.  This procedure results in  a sigma term of $\sigma_{\pi N} =(59.6\pm 7.4)$  MeV  consistent with phenomenology. Ref.~\cite{Gupta:2021ahb}  also states that further calculations are needed to confirm their result and finally resolve the tension between lattice QCD and phenomenology. I also note that  pionic atom data are used in  \cite{Hoferichter:2015dsa} and there is possibly an additional  uncertainty due to medium effects that are not included in the analyses.

The strangeness sigma term \cite{Aoki:2021kgd} is presented  in Fig.~2.  The result for $N_f=2+1+1$ is $\sigma_s=41.0(8.8) \rm MeV$, and for $N_f=2+1$, $\sigma_s=52.9(7.0)$ MeV. There is no need for a very large value of $\sigma_s$.
\begin{figure}[t]\label{SIGMAS}
 \begin{center}
 \includegraphics[width=.516\textwidth]{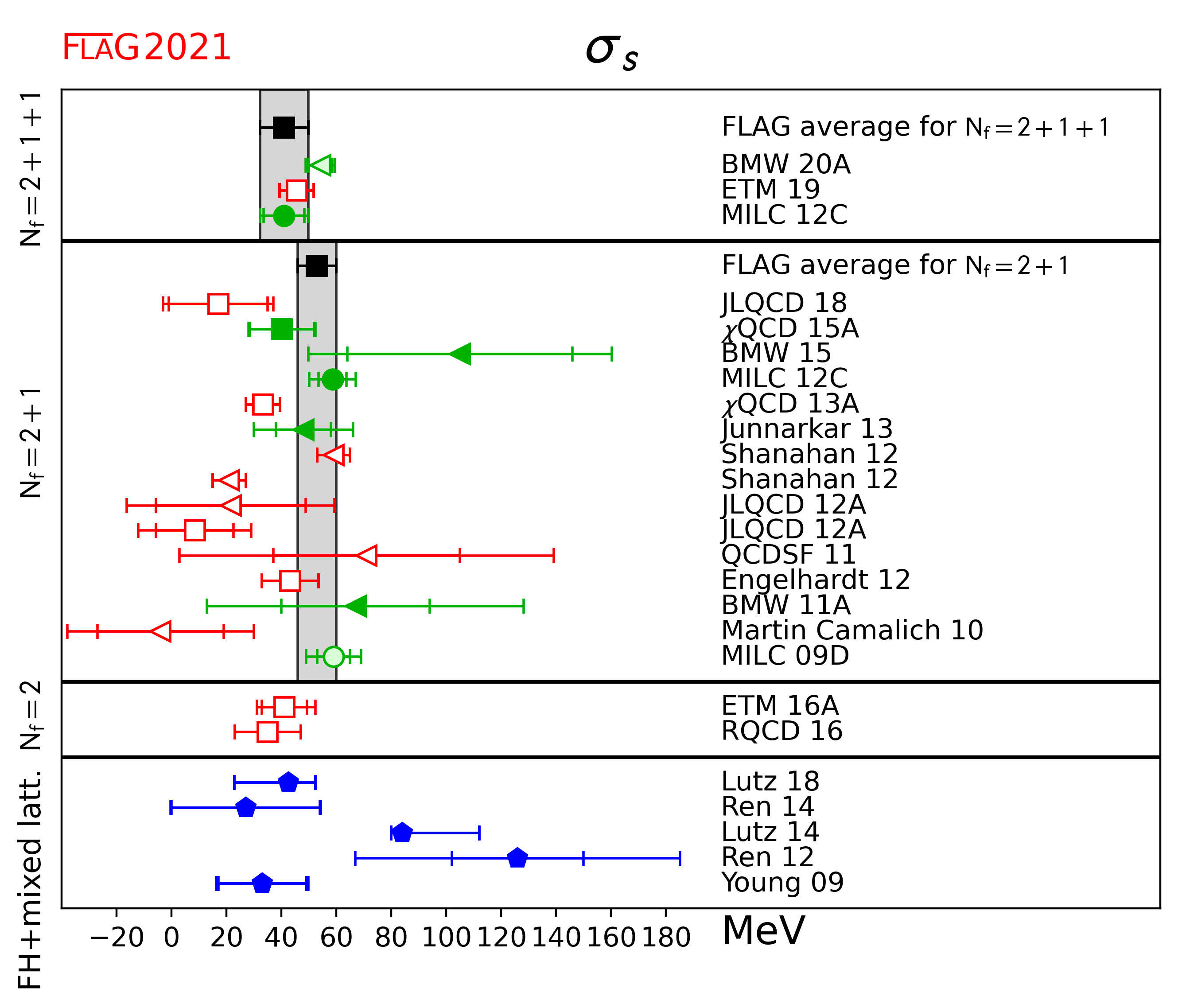} 
 \caption{ Lattice results and FLAG averages for  $\sigma_s$  for the $N_f  =2, 2 + 1$, and 2 + 1 + 1 flavor calculations. Determinations via the direct approach are indicated by squares and the Feynman-Hellmann method by triangles. Results from calculations which analyze more than one lattice data set within the Feynman-Hellmann approach \cite{Young:2009zb,Ren:2012aj,Lutz:2014oxa,Ren:2014vea,Lutz:2018cqo}, 
 are shown for comparison (pentagons)  
 (circles). he results that pass the FLAG criteria are shown in green, and are reasonably consistent. }
 \end{center}
\end{figure} 

 \section{Momentum dependence of the  low-energy $\pi-N$ phase shifts}
 
  A nice discussion of these phase shifts is presented in the book~\cite{Ericson:1988gk}. This is updated here  based on more recent data from the SAID data base using the WI08 fit \cite{Workman:2012hx} which is an update of that of ~\cite{Arndt:2006bf}.
  
 For $s$-wave scattering the isospin $1/2$ phase shifts are designated as $S_{11}$, while those of isospin $3/2$ are termed $S_{33}$. These are shown in Fig.~3. The range of pion lab momenta is from 0 to 350 MeV/c which corresponds to a range in lab kinetic energy from 0 to 240 MeV, and  cm momenta, c from 0 to 260 MeV/c.
 

 These  phase shifts of small magnitude  are  approximately linear in   the cm  momentum $q=M/\sqrt{s} q_L$ for this 
 values of $q$ up to about  $q_{\rm cm} $ of about 100 MeV/c. There $\d(S_{11}\approx {4.9\times 10^{-2}}\, q/\rm (MeV/c) deg$ and
 $\d(S_{31})\approx {- 2.9 \times 10^{-2}} q/\rm (MeV/c)deg$. The combination $\d(S_{11})+2\d(S_{31})$ nearly vanishes, consistent with early chiral perturbation theory. 
   Effective range corrections must be included to describe the phase shifts for momenta  greater than about 
  90 MeV/c. Detailed chiral perturbation theory   calculations are presented in
  Refs.~\cite{Fettes:1998ud,Fettes:1998wf,Fettes:2000gb,Fettes:2000xg,Fettes:2000vm,Fettes:2001cr}.
These have been updated in Refs.~\cite{Ditsche:2012fv,Hoferichter:2015hva,Hoferichter:2012wf,Baru:2010x,Baru:2011bw}.

\begin{figure}[h]\label{S}
 \begin{center}
 \includegraphics[width=.5\textwidth]{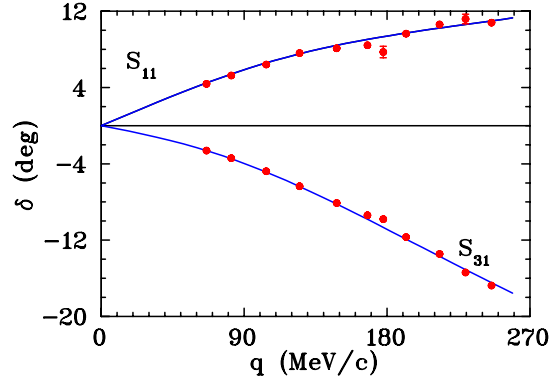} 
 \caption{$S_{11}$ and $S_{33}$ phase shifts as a function of the cm momentum of the pion beam, $q$. The points are from the single-energy fits and the curves are energy dependent fits. The data are from the SAID data base, \href{https://gwdac.phys.gwu.edu/}{SAID} and are from  the WI08 fit \cite{Workman:2012hx}, an update of that of ~\cite{Arndt:2006bf}. }
 \end{center}
\end{figure} 

\begin{figure}[h] 
 \begin{center}
 \includegraphics[width=.5\textwidth]{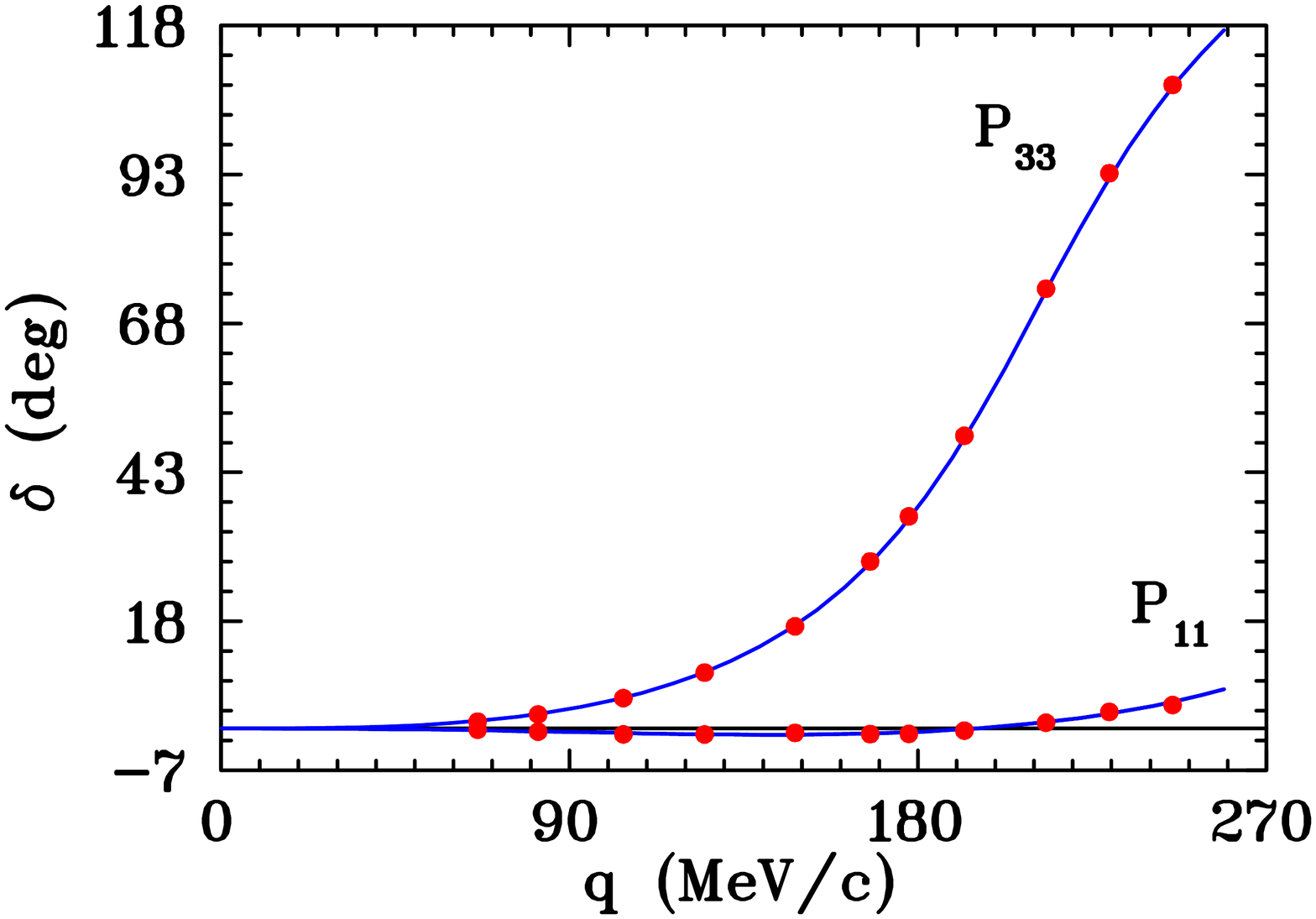} 
 \caption{$P_{33}$ and $P_{11}$ phase shifts as a function of the cm  momentum of the pion, $q$. The points are from the single-energy fits and the curves are energy dependent fits. The data are from the SAID data base, \href{https://gwdac.phys.gwu.edu/}{SAID} and are from  the WI08 fit \cite{Workman:2012hx}, an update of that of ~\cite{Arndt:2006bf}. }
 \label{P33P11}
 \end{center}
\end{figure}

The $p-$wave phase shifts are designated $P_{2I,2J}$. The $P_{33}$ and $P_{11}$  phase shifts are shown in Fig.~\ref{P33P11}. They are proportional to $q^3$ near threshold.
The $P_{33}$ channel is clearly dominant, with the phase shift rising  rapidly  to $90^\circ$ at $q\approx 292\,\rm MeV/c$, the position of the $\Delta$ resonance. The $P_{11}$  phase shift is small at the low energies of relevance here. At low energies it is negative and crosses 0 at about $q=195$ MeV/c and  rises at higher energies, reflecting the broad $N^*$(1440) resonance.

\begin{figure}[h] 
 \begin{center}
 \includegraphics[width=.5\textwidth]{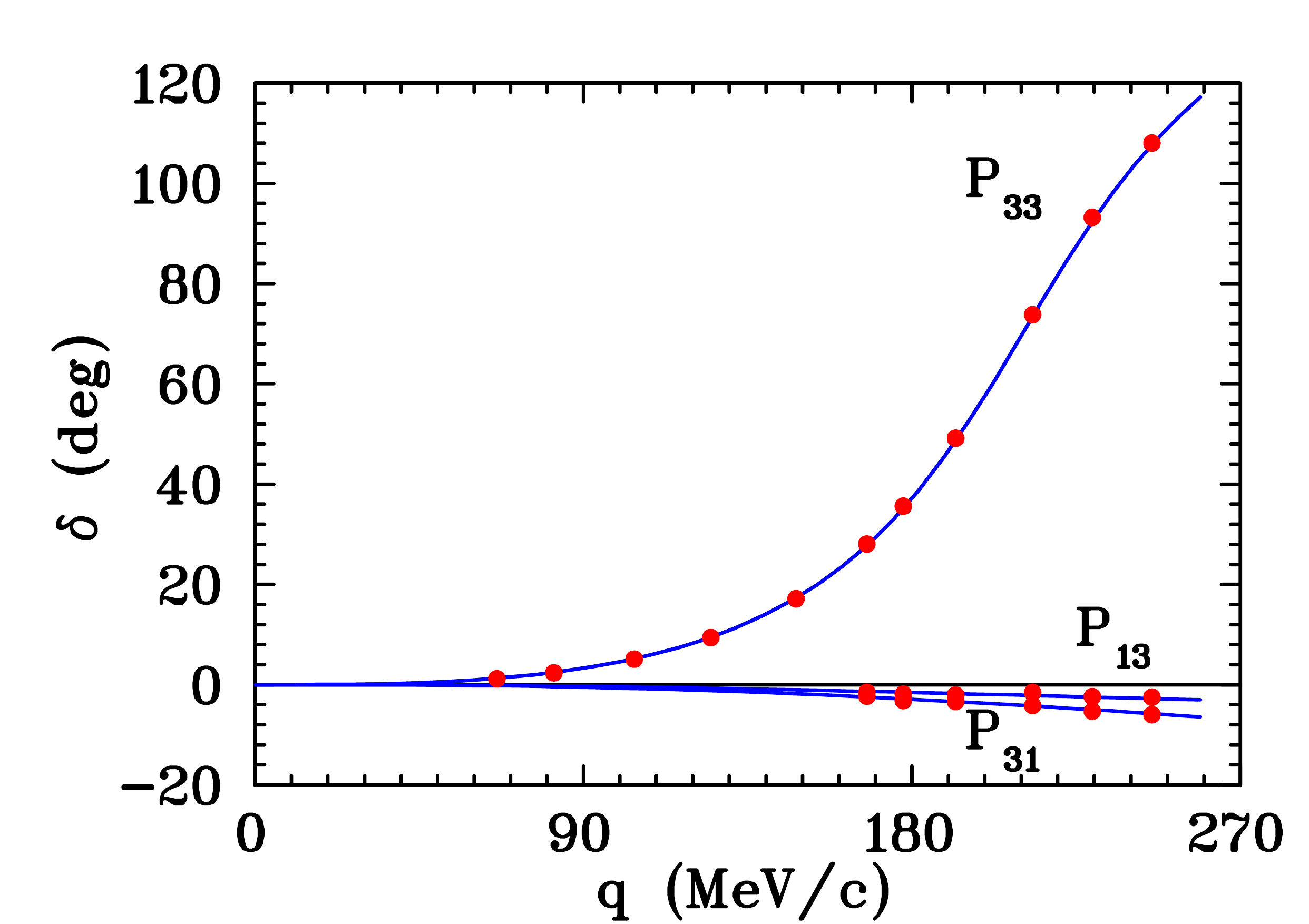} 
 \caption{ $P_{33}$ and $P_{11}$ phase shifts as a function of the cm  momentum of the pion, $q$. The points are from the single-energy fits and the curves are energy dependent fits. The data are from the SAID data base, \href{https://gwdac.phys.gwu.edu/}{SAID} and are from  the WI08 fit \cite{Workman:2012hx}, an update of that of ~\cite{Arndt:2006bf}. }
 \label{P33P13}
 \end{center}
\end{figure}

The $P_{33},\, P_{13} $ and $P_{31}$  phase shifts are shown in Fig.\ref{P33P13}..
The  other phase shifts remain small in the low-energy range of interest.

The most detailed treatment of low-energy phase shifts is to be found in Refs.~\cite{Ditsche:2012fv,Hoferichter:2015hva,Hoferichter:2012wf,Baru:2010xn,Baru:2011bw}.
These authors found that much higher precision (and accuracy) can be achieved by combining dispersion relations with the experimental input that is available (either at threshold from scattering lengths or at higher energies from phase-shift analyses). The low-energy data set for $\pi N$ scattering is scarce.  So Ref.~\cite{Hoferichter:2015hva}  summarizes  that one can do much better by combining theory input in the form of Roy-Steiner equations, experiment for the scattering lengths from pionic atoms, and the phase-shift analyses only for the high-energy input. 

The figures in this section seem to indicate that the data set is in good shape. However, as pointed out in Sect.~III, improved knowledge of 
$\s_{\pi N}$ requires better data. Section.~VI below also points out that understanding of isospin violation requires better low-energy data.
The Muon Scattering Experiment (MUSE) \cite{MUSE:2017dod,Cline:2021vlw,Cline:2021ehf} intends to obtain a better determination of  the proton radius, and more generally test lepton-universality by comparing $\m^\pm p$ with $e^\pm p$ scattering. Pion-nucleon scattering is an important background so the collaboration plans to measure  $\pi^\pm$-nucleon  cross sections at  beam momenta of  about 115, 160, and 210 MeV/c. The angular range is 20 - 100 degrees in the lab. These measurements would greatly improve the data base.

\section{analysis of the $\Delta_{33}$  resonance}
\def\D{\Delta}\def\g{\gamma}\def\o{\omega}
The $\Delta_{33}$  resonance  (or $\D$) is the simplest of all resonances. Nevertheless it has some interesting complications caused by the need to include the crossed nucleon Born graph and the effects of renormalizing the $\D$ mass caused by the need to maintain unitarity.  The aim here is to present a simple discussion to illustrate the features that are perhaps the minimal ones needed to understand more complicated resonances.
These basic features were first included in the Cloudy Bag Model calculations~\cite{Miller:1979kg,Theberge:1980ye} and are absent in the calculations of \cite{Oset:1981ih,Cheung:1983rz}.
 See Ref.~\cite{Long:2009wq} for a   detailed chiral perturbation theory analysis.

\begin{figure}[h] 
 \begin{center}
 \includegraphics[width=.5\textwidth]{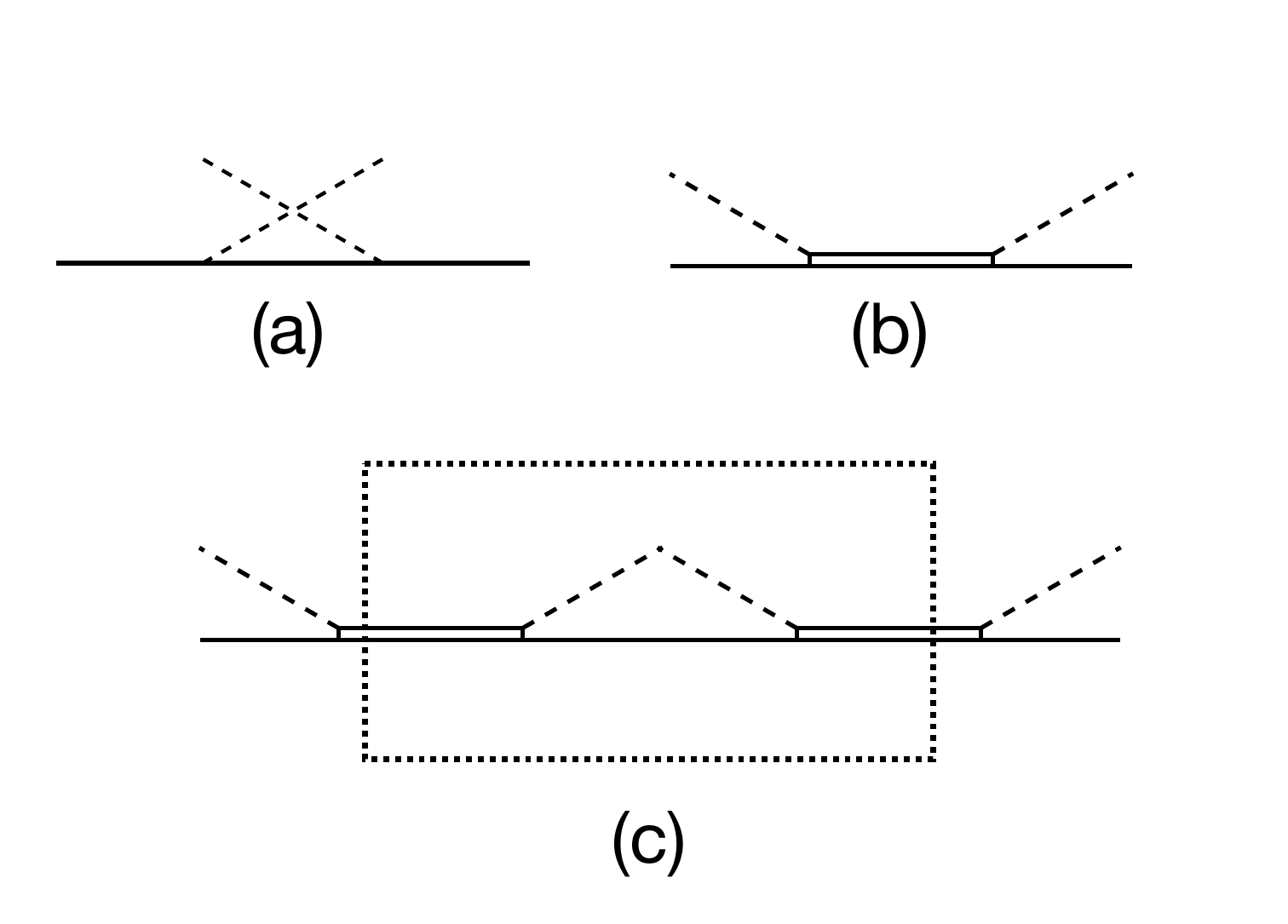} 
 \caption{Irreducible diagrams in the 33 channel. (a) Crossed nucleon term (b) Excitation of bare $\D$. (c) Iteration of (b) via an integral equation}
 \label{Born}
 \end{center}
\end{figure} 

The starting point is the Born graphs  (irreducible diagrams)  of Fig.~\ref{Born} a and b. The iteration of $\D$ excitation term, Fig. 6 c,  via an integral equation causes the inclusion of terms (isolated in the box) that correspond to the dressing of the $\D$ and a shift in the resonance peak away from the energy of the bare $\D$, $M_\D^0$. The iteration of the crossed nucleon term is not shown in Fig.~\ref{Born} but must be included.  The crossed graph with an intermediate $\D$ is not included for simplicity and because it is small.

Detailed discussion of the methods to project amplitudes onto the 33 channel are shown in the book~\cite{Ericson:1988gk}. Here I use projected amplitudes in which the (cm) Born terms  take the form
\bea \hat V(s)=\g(s) |\D\rangle\langle\D| \eea
with $s$ the square of the $\pi N$  cm energy,
\bea &\g(s)={1\over 3}{M\over \sqrt{s}}{q^2}
\left[{f_{\pi N}^2\over 4\pi}{2M\over 2 E(s)\o(s) }
+{f_{\pi N\D}^2\over 4\pi} \frac{2M_\D^0}{s-{M_\D^0}^2}\right]\nonumber\\&
\eea
and $|\D\rangle$ representing the form factor arising from the finite extent of the pion and baryon. The first term in the bracket
is inversely proportional to $M^2-\bar u$, in which $\bar u $ accurately approximates the Mandelstam variable $u$~\cite{Ericson:1988gk}. I use $\langle q|\D\rangle= 1/(1+q^2/M_A^2)^2\equiv v(q)$ with $M_A$ the axial mass of 1030 MeV. I use $f_{\pi N}^2/(4\pi)=0.08 $ and $f_{\pi N\D}=1.5 f_{\pi N}$, obtained from the large $N_c $ limit.
  See Refs.~\cite{Siemens:2016jwj,PhysRevD.87.054032,NavarroPerez:2014ovp,Epelbaum:2008ga}. 
  
The $K$- matrix then takes the form
\bea \hat K(s)= D(s)  |\D\rangle\langle\D| .\eea 
The on-shell $K-$matrix element in a given channel is $\tan\d(q)/q$, with $q$ the cm momentum.
Solving the integral equation:
\bea \hat K(s)=\hat V(s) +\hat V(s)P G(s) \hat K(s),\eea
with $P$ representing a principle-value integral,
yields the result
\bea D(s)={\g(s)\over 1- \g(s) v^2(q) \langle \D|G(s)|\D \rangle}\eea
with Green's function matrix element given by:
\bea  \langle \D|G(s) |\D\rangle=P\int dp\, {p^4 v^2(p)}\,  {1\over s-(E(p)+\o(p))^2}.
\eea
Defining the smooth function $\hat\g(s)\equiv (s-{M_\D^0}^2)\g(s) $ allows $D(s) $ to be re-written as
\bea D(s)={\hat \g(s)\over s-{M_\D^0}^2 - \hat\g(s) v^2(q) \langle \D|G(s)|\D \rangle}\eea
This shows that the resonance position is shifted away from $\sqrt{s}=M_\D^0$ by the iteration of both the nucleon crossed Born term and the $\D$ excitation.  The calculations of~\cite{Oset:1981ih,Cheung:1983rz} use $\hat K(s)\to  \g(s)  |\D\rangle\langle\D|$, a procedure that amounts to unitarizing the Born approximation..

\begin{figure}[h] 
 \begin{center}
 \includegraphics[width=.5\textwidth]{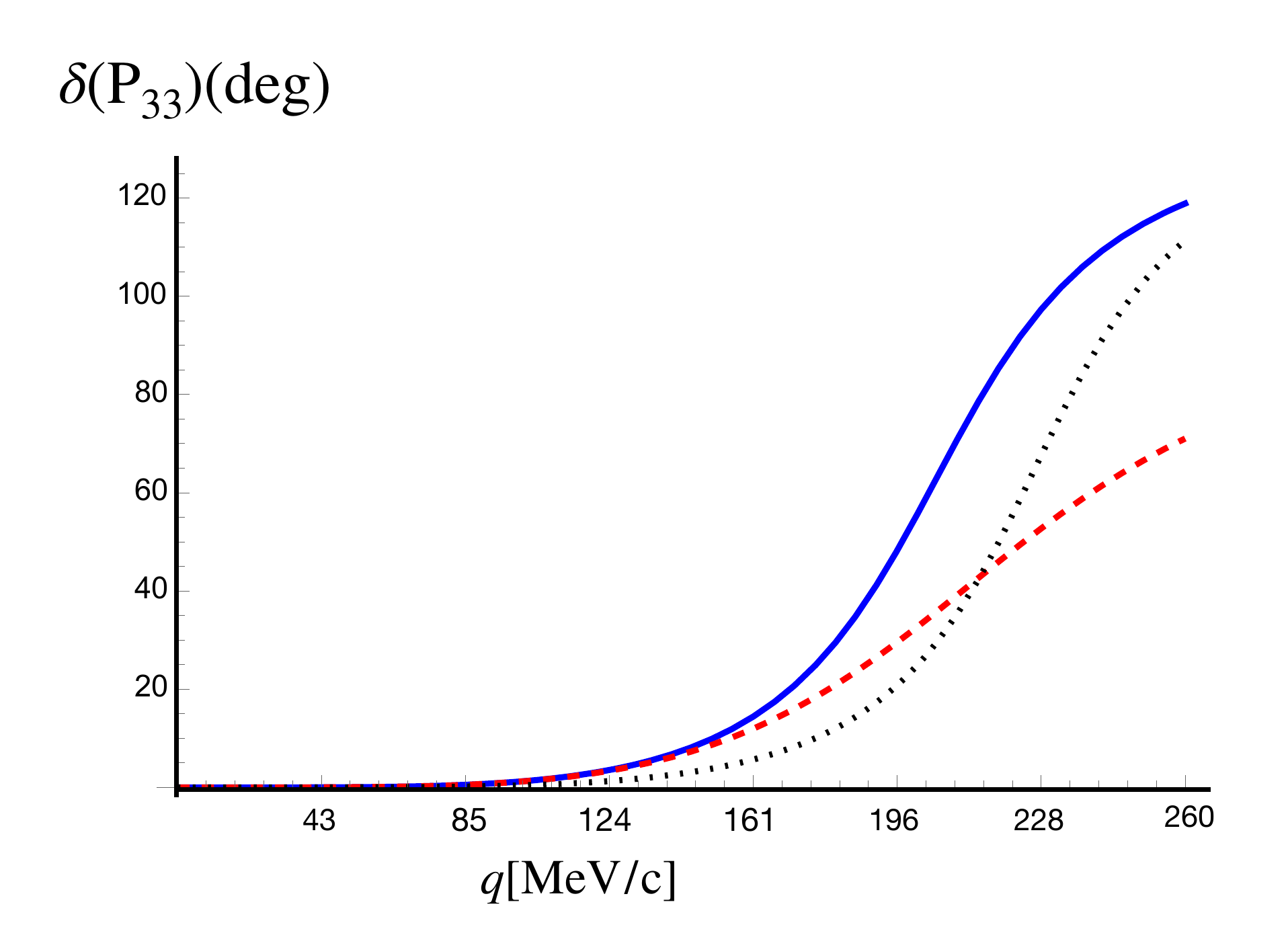} 
 \caption{$P_{33}$ phase shifts computed with the model explained in the text.  Solid (blue) curve: full calculation. Dashed(red): Unitarized Born approximation of~\cite{Oset:1981ih,Cheung:1983rz} . Dotted: full calculation without the crossed nucleon Born term of fig (a).}
 \label{delta33}
 \end{center}
\end{figure}

Results of a numerical calculation are shown in Fig.~\ref{delta33}. The solid (blue) curve is the full calculation that approximately reproduces the data. The dashed (red) curve shows result of using the Born approximation procedure of  \cite{Oset:1981ih,Cheung:1983rz} with the present parameters. The dotted (black) curve shows the results obtained by omitting the crossed nucleon Born term. The figure shows that  the cross Born term must be kept and that the Born approximation to the $K$-matrix is a poor approximation.

 \section{Isospin violation in pion-nucleon interactions}
 The basis for isospin violation in the strong interaction is the mass difference between the up and down quarks, as written the $A_3$ term of \eq{mud}. All isospin violations within the standard model arise from this term and electro-weak interactions,  as stressed in Ref.~\cite{Miller:1990iz}. 
 
 Weinberg~\cite{Weinberg:1994tu}
 wrote the effective interaction arising from this term as a charge symmetry breaking (csb)  contribution to the effective interaction
 \bea \delta{\cal L}_{\rm csb}= -B\left[\bar \psi t_3\psi-{1\over 2 f_\pi^	2}\left({\pi_3\over 1+{\bfpi^2\over 4f_\pi^2}}\right)\bar \psi {\bf t}
 \cdot\bfpi\psi\right],	\label{full} \eea
with $B\propto m_u-m_d$.
This term leads to a term in the scattering length for the process of \eq{pin}:
\bea \d a_{ba}({\rm csb})={1\over 4\pi(1+m_\pi/m_N)}{B\over 2f_\pi^2}(t_a\d_{3b}+t_b\d_{3a}).
\label{acsb}
\eea
Weinberg comments that the immediate consequence of \eq{full} is that  ``isospin violation never appears in a process that does not involve at least one neutral pion.".  I believe that Weinberg meant to say that  ``isospin violation from the strong interaction never appears in a process that does not involve at least one neutral pion." 	For example, the Coulomb interaction causes isospin violating differences between $\pi^+ d$ and $\pi^- d$ scattering.

Weinberg \cite{Weinberg:1977hb} used \eq{acsb} 
 estimated that the $\pi^0n\to\pi^0n$ scattering length was 30\% larger than the one for $\pi^0p\to\pi^0p$, although both are of the order of 0.01 fm. Of course, the related scattering experiments are not possible with current techniques.
 
I turn to the more general case of $\pi N$ scattering. 
If isospin invariance is respected one can write the $\pi-N$ scattering amplitude as an operator in a schematic form
\bea  f=  f_0+ f_1{\bf t}\cdot {\bf t}_\pi,\label{i}\eea
where $ f_{0,1}$ are operators that depend on relative momentum and nucleon spin. There are eight reactions of the form  $\pi ^a (q) N(p) \to  \pi^b (q') N(p') $, but only three are available to experiments \cite{Gibbs:1995dm}: $\pi^\pm p$ elastic scattering with amplitudes $f_{\pm}$ and $\pi^-p\to \pi^0 n$ charge exchange (CEX) scattering with amplitude $f_{rm CEX}$. The use of \eq{i} yields the so-called triangle identity:
\bea f_{\rm CEX}={1\over \sqrt{2}}(f_+-f_-).
\eea
The failure of this equality is a failure of isospin invariance, but not of charge symmetry. The $B$ term of \eq{full} causes no difference between the left and right-hand sides of the above equation.  Thus a powerful conclusion results: the only Standard Model effects that break the triangle identity come from  electroweak and  hadronic mass-difference effects. The neutron-proton mass difference does  arises from light quark mass differences and from the electroweak interaction, but is an indirect effect through its kinematic influence. 

If isospin symmetry is broken
the term $ f_1{\bf t}\cdot {\bf t}_\pi$ is replaced by the isospin-breaking (ib) term $f_1({\rm ib})$ with
\bea
f_1({\rm ib}) =\alpha t_z t_{\pi z} + \beta (t_xt_{\pi x}+ t_y t_{\pi y} )
\label{ib}
,\eea
with $\alpha\ne \beta$ and $\alpha-\beta\propto m_u-m_d$ if electromagnetic effects are neglected. The $\alpha$ term causes differences between $\pi^+p$ and $\pi^-p$ scattering, the terms proportional to $\beta$  are responsible for charge exchange.
One can get a quick glimpse of the size the effects by examining the kinematics of the $\pi^-p\to\pi^0n$ reaction. The $\pi^-$ experiences a Coulomb attraction of about $\a/r_p\approx 1.7$ MeV and the proton is lighter than the neutron by about 1.3 MeV the sum of these two effects of 3 MeV is offset by the $\pi^--\pi^0$  mass difference of 4.6 MeV.

 Early   analyses \cite{Gibbs:1995dm,Matsinos:1997pb} of the triangle identity in   $\pi N$ low-energy  scattering data  found  indications for significant 
 6-7 \% csb  effects in the strong-interaction sector. Fettes and Meissner~\cite{Fettes:2000vm} 
have found a smaller violation using  a third-order chiral perturbation theory calculation without electromagnetic effects, and even smaller violations once the latter were included~\cite{Fettes:2001cr}. They found -0.7\% violation in the $s$-wave. An interesting feature of this work is the inclusion of the effects of  the $\g n$ channel in the final state. A detailed  
comprehensive partial-wave
analysis of $\pi^\pm p$  elastic scattering and charge- exchange data, covering the region from threshold to 2.1 GeV in the lab pion kinetic energy found no compelling evidence for sizable isospin breaking  effects beyond those of the standard model.

Isospin breaking  through quark mass differences and virtual photons in scattering lengths was analyzed using covariant baryon chiral perturbation theory~\cite{Hoferichter:2009ez,Hoferichter:2009gn}. This work found that the so-called triangle relation that vanishes in
the isospin limit is violated by about 1.5\% consistent with earlier
findings and inconsistent with the much larger deviations extracted from the data at lowest pion momenta in~\cite{Gibbs:1995dm,Matsinos:1997pb}. Isospin violation in the triangle identity  was studied later in~\cite{RuizdeElvira:2017stg} using 
 cross-section data.  The conclusion was that  these data are simply not precise enough to test the triangle relation at the percent level.

An experimental  study of isospin symmetry in  the $\pi^-p\to\pi^0n$ reaction at momenta between 104 and 143 MeV/c was made at TRIUMF~\cite{Jia:2008rt}. This kinematic region exhibits a vanishing zero-degree cross section from destructive interference between $s$ and $p$ waves, thus yielding special sensitivity to pion-nucleon dynamics. Their data and previous data \cite{Fitzgerald:1986fg} disagree.
The key feature is that the predictions of Gibbs {\it et al.} \cite{Gibbs:1995dm,Gibbs:2004jf} and Matsinos {\it et al.} \cite{Matsinos:1997pb,Matsinos:2006sw} based on fitting $\pi^\pm p$ data that constrain the term $\a$ of \eq{ib} agree with the angular distributions obtained in ~\cite{Jia:2008rt}, but not with those obtained in \cite{Fitzgerald:1986fg}. This means that no significant isospin breaking effect is observed. The difference in data sets was attributed to normalization uncertainties and background subtraction.

The net result regarding tests of the triangle identity   is that there isn't any evidence for these huge isospin-breaking effects.

One also studies charge symmetry breaking and isospin violation involving light nuclei $\rm d, ^3H,^3He,^4He$ as detailed in the review~\cite{Miller:1990iz}. Here I focus on later experiments involving $\pi^\pm$ scattering on the deuteron\cite{Baru:1999ns}, $^3$He and $^3$H~\cite{Kudryavtsev:2001fe}.

Charge symmetry breaking effects in experimental measurements have been expressed in terms of $A_\pi$ for the deuteron:
\bea A_\pi=\frac{d\s/d\O(\pi^-d)-d\s/d\O(\pi^+d)}{d\s/d\O(\pi^-d)+d\s/d\O(\pi^+d)},
\eea
and for the nuclei with $A=3$:
\bea& r_1=\frac{d\s/d\O(\pi^+{^3}\rm H)}{d\s/d\O(\pi^-{^3}\rm He)}\\&r_2=\frac{d\s/d\O(\pi^-{^3}\rm H)}{d\s/d\O(\pi^+{^3}\rm He)}\\
&R=r_1r_2
\eea
Charge symmetry predicts that $A_\pi=1, r_1=1,r_2=1$ and $R=1$. Of course it is necessary to remove the effects of the Coulomb interaction. These experiments were performed at energies near and on that of the $\D$ resonance and one must account for the mass and width differences amongst the four charge states of the $\D$ as well as the neutron-proton mass difference and the $^3$He-$^3$H mass difference.

The early history suggested evidence for a small effect, $A_\pi\approxeq 2\%$ at 143 MeV~\cite{Masterson:1982ki}. A sizable effect was reported in the $^3$He$^3$H case. For example, $r_2=0.7\pm0.1$ for $T_\pi=256$ MeV at $\theta=82^\circ$~\cite{Dhuga:1996cg}.

Ref.~\cite{Baru:1999ns} described the charge symmetry breaking effect on the deuteron in terms of single and double scattering amplitudes, taking the $\D$ mass splitting into account. The results of that study indicate that the $\D$ mass and width differences, can along with the Coulomb interaction are of the same order as the deviations between $A_\pi $ and unity. 

Ref.~\cite{Kudryavtsev:2001fe}  described pion scattering from the three-nucleon system in terms of single- and double-scattering amplitudes. Coulomb interactions and $\D$-mass splitting are taken into account as sources of charge symmetry breaking. Reasonable agreement between  theoretical calculations and the experimental data was obtained for $T_\pi=180, 220$, and 256 MeV. For these energies, it is found that the $\D$-mass splitting and the differences in wave functions of the $A=3$ nuclei  are the most important contributions for charge symmetry breaking.  Data for $r_2$  and R at $T_\pi=$ 142 MeV did not agree with the predictions of the model, which may indicate that there are additional mechanisms for charge symmetry breaking that  are important only at lower energies.

The summary of the activity for elastic pion scattering from nuclei with $A=2$ or 3 is that the $\D$ mass differences and Coulomb effects account for most  of the charge symmetry breaking effects. Other mechanisms involving non-resonant charge symmetry breaking in $s-$wave $\pi^0$-nucleon scattering that occurs in the intermediate state, as indicated in \eq{ib} and \eq{acsb}  were not included in the calculations.

Weinberg's statements~\cite{Weinberg:1994tu} about charge symmetry breaking involving the $\pi^0$ were borne out in experiments on the $np\to \pi^0 d$ reaction~\cite{Opper:2003sb} in which charge symmetry breaking predicts a symmetry in the angular distribution about a production angle of 90$^\circ$ and in the $\rm dd\to ^4He \,\pi^0$ reaction \cite{Stephenson:2003dv,Bacher:2019tpm} that is forbidden by charge symmetry.  Ref. ~\cite{Opper:2003sb} observed a forward-backward asymmetry $[17.2 \pm8.0\rm(stat)\pm 5.5 (syst)] \times 10^{-4}$ at an incident neutron energy of 279.5 MeV. This result could be accounted for by the effects of the mass difference between up and down quarks. Refs.~\cite{Stephenson:2003dv,Bacher:2019tpm}  measured total cross sections for neutral pion production of $12.7\pm 2.2 $ pb at 228.5 MeV and $15.1\pm3.1$ pb at 231.8 MeV. The uncertainty is dominated by statistical errors. These cross sections arise fundamentally from the down-up quark mass difference and quark electromagnetic effects that contribute in part through meson mixing (e.g.,$\pi^0-\eta$) mechanisms.

The results of these experiments and the related theory are discussed in the review~\cite{Miller:2006tv}. The simplest summary is that the effects of the mass difference between up and down quarks of \eq{mud} cause the charge symmetry breaking . Initial calculations of the manifestation of the mass difference in terms of the necessary hadronic degrees of freedom accounted for the findings of  experiments with reasonable accuracy. The effects of the light-quark mass difference has made its presence felt in nuclear physics.

\section{Summary}

Chiral effective field theory is a greatly useful tool to  understand low-energy pion-nucleon interactions and their connection with the fundamental theory of QCD. The breaking of chiral symmetry via quark mass differences is manifest in specific features  of Lagrangians using hadronic interactions. The  progress in understanding the sigma term and its widespread importance is reviewed. 
The extraction of the sigma term would benefit from improving the low-energy data set.
The  features of the low-energy $\pi$N phase shifts are reviewed. The complicated physics of the simplest baryon resonant state--the $\D$ is explained. Recent progress in understanding isospin violation in theory and fact is reviewed.
 The experimental observations of isospin violation seem to be accounted for by the effects of light-quark mass differences and electromagnetic effects. However,  higher energy phase shifts provide the dominant experimental information. This is a second reason for why it would be very useful to have better low-energy pion-nucleon scattering data.
 
The study of pion-nucleon scattering is equivalent to the study of the fundamental strong interaction of QCD. Its importance remains vital to this day. 
 
 \section*{Acknowledgements} This work   was supported by the US Department of Energy Office of Science, Office of Basic Energy Sciences program under Grant No. DE-FG02-97ER- 41014. I thank Avraham Gal, Ronald Gilman, Andre Walker-Loud and Martin Hoferichter for  useful discussions and Igor Strakovsky for providing Figs. 3-5.

\end{document}